\documentclass[aps,prl,preprint,tightenlines,superscriptaddress,showpacs]{revtex4}
\def\mystyle{4}

\if\mystyle4
  \def\bellelogo{\vbox to 16mm{
                 \vss\hbox{\resizebox{!}{3cm}{
                 \includegraphics{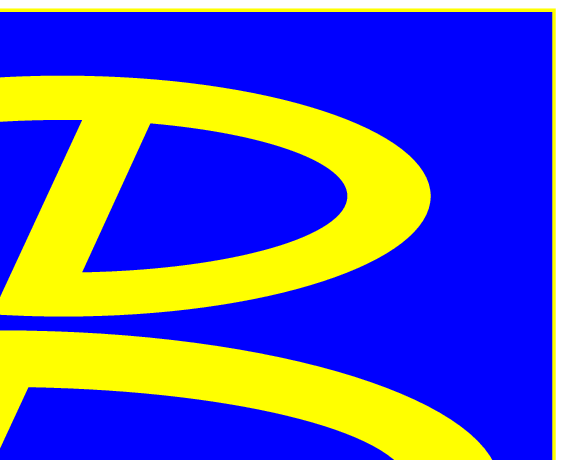}}}}\vspace{-1cm}}
  \def\preprintA{\hbox{\hfil Belle Preprint 2006-09}}
  \def\preprintB{\hbox{\hfil KEK Preprint 2005-106}}
 \def\preprintC{}
  \def\SomeSpaceIfPreprint{\quad\\[1cm] \Large}
\else
 \def\bellelogo{}
 \def\SomeSpaceIfPreprint{}
 \def\preprintA{}
 \def\preprintB{}
 \def\preprintC{}
\fi

\if\mystyle2
  \def\mydate{\date{Mar. 8, 2006, Version 0.994, option 4}}
\else
  \def\mydate{\date{Mar. 8, 2006}}
\fi

\usepackage{graphicx}

\def\elel{\ell^+\ell^-}

\def\KP{K^+}
\def\piP{\pi^+}
\def\piM{\pi^-}
\def\piZ{\pi^0}

\def\KS{K^0_S}

\def\Kst{K^*}
\def\KstZ{K^{*0}}
\def\KstP{K^{*+}}

\def\Kstll{\Kst\elel}
\def\Kstlh{\Kst\ell^{\pm} h^{\mp}}
\def\Ksthh{\Kst h^+ h^-}

\def\Kstarlh{\Kst\ell h}
\def\Kstarhh{\Kst h h}

\def\KorKstarll{K^{(*)}\elel}

\def\Xs{X_s}
\def\Xsll{\Xs\ell^+\ell^-}

\def\BtoKstarll{B\to K^*\ell^+\ell^-}

\def\fbinv{{\rm~fb}^{-1}}

\def\Mbc{M_{\rm bc}}

\def\sig{{\rm sig}}
\def\Cseven{\tilde{C}_7^{\rm eff}}
\def\Cnine{\tilde{C}_9^{\rm eff}}
\def\Cten{\tilde{C}_{10}^{\rm eff}}
\def\Ci{\tilde{C}_i}

\def\PM#1#2{\,^{+#1}_{-#2}}

\def\Journal#1#2#3#4{{#1} {\bf #2}, #3 (#4)} 

\def\NPB{Nucl. Phys. B}
\def\PLB{Phys. Lett. B}
\def\PRL{Phys. Rev. Lett.}
\def\PRD{Phys. Rev. D}
\def\ZPC{Z. Phys. C}

\def\RMP{Rev. Mod. Phys.}

\def\be{\begin{equation}}
\def\ee{\end{equation}}
\def\ba{\begin{eqnarray}}
\def\ea{\end{eqnarray}}

\begin{document}


\bellelogo

\preprint{\vbox{
  \preprintA
  \preprintB
  \preprintC
}}

\title{\SomeSpaceIfPreprint
Measurement of Forward-Backward Asymmetry and Wilson Coefficients in \boldmath$\BtoKstarll$}

\affiliation{Budker Institute of Nuclear Physics, Novosibirsk}
\affiliation{Chiba University, Chiba}
\affiliation{Chonnam National University, Kwangju}
\affiliation{University of Cincinnati, Cincinnati, Ohio 45221}
\affiliation{University of Hawaii, Honolulu, Hawaii 96822}
\affiliation{High Energy Accelerator Research Organization (KEK), Tsukuba}
\affiliation{Institute of High Energy Physics, Chinese Academy of Sciences, Beijing}
\affiliation{Institute of High Energy Physics, Vienna}
\affiliation{Institute of High Energy Physics, Protvino}
\affiliation{Institute for Theoretical and Experimental Physics, Moscow}
\affiliation{J. Stefan Institute, Ljubljana}
\affiliation{Kanagawa University, Yokohama}
\affiliation{Korea University, Seoul}
\affiliation{Swiss Federal Institute of Technology of Lausanne, EPFL, Lausanne}
\affiliation{University of Ljubljana, Ljubljana}
\affiliation{University of Maribor, Maribor}
\affiliation{University of Melbourne, Victoria}
\affiliation{Nagoya University, Nagoya}
\affiliation{Nara Women's University, Nara}
\affiliation{National Central University, Chung-li}
\affiliation{National United University, Miao Li}
\affiliation{Department of Physics, National Taiwan University, Taipei}
\affiliation{H. Niewodniczanski Institute of Nuclear Physics, Krakow}
\affiliation{Nippon Dental University, Niigata}
\affiliation{Niigata University, Niigata}
\affiliation{Nova Gorica Polytechnic, Nova Gorica}
\affiliation{Osaka City University, Osaka}
\affiliation{Osaka University, Osaka}
\affiliation{Panjab University, Chandigarh}
\affiliation{Peking University, Beijing}
\affiliation{Princeton University, Princeton, New Jersey 08544}
\affiliation{RIKEN BNL Research Center, Upton, New York 11973}
\affiliation{Saga University, Saga}
\affiliation{University of Science and Technology of China, Hefei}
\affiliation{Seoul National University, Seoul}
\affiliation{Shinshu University, Nagano}
\affiliation{Sungkyunkwan University, Suwon}
\affiliation{University of Sydney, Sydney NSW}
\affiliation{Tata Institute of Fundamental Research, Bombay}
\affiliation{Toho University, Funabashi}
\affiliation{Tohoku Gakuin University, Tagajo}
\affiliation{Tohoku University, Sendai}
\affiliation{Department of Physics, University of Tokyo, Tokyo}
\affiliation{Tokyo Institute of Technology, Tokyo}
\affiliation{Tokyo Metropolitan University, Tokyo}
\affiliation{Tokyo University of Agriculture and Technology, Tokyo}
\affiliation{University of Tsukuba, Tsukuba}
\affiliation{Virginia Polytechnic Institute and State University, Blacksburg, Virginia 24061}
\affiliation{Yonsei University, Seoul}
   \author{A.~Ishikawa}\affiliation{Department of Physics, University of Tokyo, Tokyo} 
   \author{K.~Abe}\affiliation{High Energy Accelerator Research Organization (KEK), Tsukuba} 
   \author{I.~Adachi}\affiliation{High Energy Accelerator Research Organization (KEK), Tsukuba} 
   \author{H.~Aihara}\affiliation{Department of Physics, University of Tokyo, Tokyo} 
   \author{D.~Anipko}\affiliation{Budker Institute of Nuclear Physics, Novosibirsk} 
   \author{Y.~Asano}\affiliation{University of Tsukuba, Tsukuba} 
   \author{T.~Aushev}\affiliation{Institute for Theoretical and Experimental Physics, Moscow} 
   \author{A.~M.~Bakich}\affiliation{University of Sydney, Sydney NSW} 
   \author{V.~Balagura}\affiliation{Institute for Theoretical and Experimental Physics, Moscow} 
   \author{M.~Barbero}\affiliation{University of Hawaii, Honolulu, Hawaii 96822} 
% ...

\collaboration{The Belle Collaboration}

\mydate

\begin{abstract} 

We report the first measurement of the forward-backward asymmetry and 
the ratios of Wilson coefficients $A_9/A_7$ and $A_{10}/A_7$ in $B \to \Kstll$,
where $\ell$ represents an electron or a muon.
We observe a large integrated forward-backward asymmetry with a significance of
3.4$\sigma$.
The results are obtained from a data sample containing 386
$\times 10^6$ $B\bar{B}$ pairs that were collected on the $\Upsilon(4S)$ resonance
with the Belle detector at the KEKB asymmetric-energy $e^+ e^-$ collider.
\end{abstract}




\maketitle


Flavor-changing neutral current $b \to s$ processes proceed 
via loop diagrams in the Standard Model (SM). If additional diagrams with
non-SM particles contribute to such processes, the decay rate and other properties are modified.
Such contributions may change the
Wilson coefficients~\cite{BBL} that parametrize the strength of the short
distance interactions. The $b \to s \ell^+ \ell^-$ amplitude is described
by the effective Wilson coefficients $\Cseven$, $\Cnine$ and 
$\Cten$, whose terms have been calculated up to next-to-next-to-leading order (NNLO)~\cite{WC} in the SM. 

The magnitude of $\Cseven$ is strongly constrained from measurements of 
$B \to X_{s}\gamma$~\cite{btosgamma,C7}
and a large area of the ($\Cnine$, $\Cten$) plane is excluded by branching fraction
measurements of $B\to K^{(*)} \ell^+\ell^-$
and $B \to \Xsll$~\cite{BelleKstarll,KllXsll,ALGH,Gambino}.
However the sign of $\Cseven$ and values of $\Cnine$ and $\Cten$ are not yet determined.
Measurement of the forward-backward asymmetry and differential decay rate as
 functions of $q^2$ and $\theta$ for $B \to \Kstll$ constrains the relative signs and magnitudes of these coefficients~\cite{C10flip,C7C9C10}. 
Here, $q^2$ is the squared invariant mass of  the dilepton system, 
and $\theta$ is the angle between the momenta of the negative (positive)
lepton and the $B$ ($\bar{B}$) meson in the dilepton rest frame.
The forward-backward asymmetry is defined using the differential decay width, $g(q^2,\theta)= d^2\Gamma/dq^2d\cos\theta$~\cite{bib:abhh}, as 
\begin{eqnarray}
{\cal A}_{\rm FB}(q^2) = { \int_{-1}^1 {\rm sgn}(\cos\theta) g(q^2,\theta)\,d\cos\theta \over \int_{-1}^1 g(q^2,\theta)\,d\cos\theta}.
\label{eq:afb}
\end{eqnarray}
The numerator in Eq.~\ref{eq:afb} does not cancel due to interference between the electroweak penguin and box diagrams, and can be expressed in terms of Wilson coefficients as
\begin{eqnarray}
& &  \int^1_{-1} {\rm sgn}(\cos\theta) g(q^2,\theta) d\cos\theta \nonumber \\
&=&  -\Cten \xi(q^2) \left( {\rm Re}(\Cnine)F_1 + {1\over q^2}\Cseven F_2 \right) ,
\end{eqnarray}
where $\xi$ is a function of $q^2$, and $F_{1,2}$ are functions of form factors (the full expression can be found in Ref.~\cite{bib:abhh}).

In this Letter, we report the first measurement of the forward-backward asymmetry
and ratios of Wilson coefficients in $\BtoKstarll$. 
We use a $357\fbinv$ data sample containing 386 $\times 10^6$ 
$B\bar{B}$ pairs taken at the $\Upsilon(4S)$ resonance.
We also study the $B^+ \to K^+ \ell^+ \ell^-$ mode, which
is expected to have very small forward-backward asymmetry even in the existence of
new physics~\cite{NoAFBKll}. Charge-conjugate modes are included throughout this Letter.

The data were taken at the KEKB
collider~\cite{KEKB} and collected
with the Belle detector~\cite{Belle}.  The detector
consists of a silicon
vertex detector, a central drift chamber, 
aerogel Cherenkov counters, time-of-flight
scintillation counters, an electromagnetic calorimeter,
and a muon identification system.


The event reconstruction procedure is the same as described in our previous
report~\cite{BelleKstarll}.  The following final states are used to reconstruct $B$ candidates:
$\KstZ \elel$, $\KstP \elel$, and $\KP \elel$,
with subdecays
$\KstZ\to\KP\piM$, $\KstP\to\KS\piP$ and $\KP\piZ$,
$\KS\to\piP\piM$, and $\piZ\to\gamma\gamma$.
Hereafter, $\KstZ \elel$ and $\KstP \elel$ are combined and called
$K^* \elel$.

 We use two variables defined in the center-of-mass (CM) frame to select $B$
candidates: the beam-energy constrained mass $M_{\rm{bc}} = \sqrt{ ({E_{\rm{beam}}^*/c^2})^2 -
({p_{B}^*/c})^{2}}$ and the energy difference
$\Delta E = E_{B}^* - E_{\rm{beam}}^* $, where 
$p_{B}^*$ and $E_B^*$ are the measured CM momentum and energy of the $B$
candidate, and $E_{\rm{beam}}^*$ is the CM beam energy. When multiple candidates are
found in an event, we select the candidate with the smallest value of $|\Delta E|$.

The dominant background consists of $B\bar{B}$ events where both $B$ mesons decay semileptonically. We suppress this using missing energy and
$\cos\theta_B^*$, where $\theta_B^*$ is the angle between the flight direction of the $B$ meson and the beam axis in the CM frame.
These quantities are combined to form signal and background likelihoods, ${\cal{L}}_{\rm{sig}}$ and
${\cal{L}}_{B\bar{B}}$, and
event selection is then performed using the ratio ${\cal{R}}_{B\bar{B}} =
{\cal{L}}_{\rm{sig}} / ({\cal{L}}_{\rm{sig}} + {\cal{L}}_{B\bar{B}})$.
The continuum ($e^+e^- \to q\bar{q}$, $q=u,d,s,c$) background is suppressed using
a likelihood ratio ${\cal{R}}_{\rm cont}$ 
(defined similarly to ${\cal{R}}_{B\bar{B}}$)
that depends on three variables;
a Fisher discriminant~\cite{fd} calculated from the energy flow in 9 cones along the $B$ candidate sphericity axis and 
the normalized second Fox-Wolfram moment~\cite{fw},
the angle between the beam axis
and the CM sphericity axis calculated with tracks used in the $B$ 
meson reconstruction, and $\cos\theta_B^*$.
Backgrounds from $B \to J/\psi X_s, \psi(2S) X_s$ decays, below referred to as $B \to \psi X_s$, are rejected using the
dilepton invariant mass.
Backgrounds from photon conversions and $\pi^{0}$ Dalitz
decays are suppressed by requiring the $e^+e^-$ invariant mass to be above 140~MeV/$c^2$. 

The signal box is defined as $|M_{\rm bc}-m_{B}|<8$~MeV/$c^{2}$ for
both lepton modes and
$-55\, (-35)\, {\rm MeV} < \Delta E < 35\, {\rm MeV}$ for the electron (muon) mode.
We optimize the selections on ${\cal{R}}_{\rm cont}$ and ${\cal{R}}_{B\bar{B}}$
for each $K^*$ decay mode and each lepton mode to maximize sensitivity to events with $q^2 <$  6~GeV${}^2$/$c^2$.


To determine the signal yield, we perform an unbinned maximum-likelihood fit to the $M_{\mathrm{bc}}$ distribution for events that lie within the $\Delta E$ signal window.
The fit function includes signal, cross-feeds and other background components. The cross-feeds are misreconstructed $\KorKstarll$ events with correct (``CF'') and incorrect (``IF'') flavor tagging. The cross-feed from $\Xsll$ events other than $\KorKstarll$ is negligible. The other backgrounds come from dilepton background, combinatorial $K^{(*)} \ell^{\pm} h^{\mp}$, $K^{(*)} h^{+} h^{-}$ and $\psi X_s$ events, where $h$ represents a pion or a kaon. The dilepton background refers to the sum of all background sources with two leptons where the lepton is from (semi)leptonic meson decays, photon conversions and $\pi^0$ Dalitz decays. The $K^{(*)} h^{+} h^{-}$ is from both combinatorial background and $B$ meson decays.

The shape for cross-feed events is parametrized by a sum of an ARGUS function~\cite{ARGUS} and a Gaussian whose parameters are determined from Monte Carlo (MC) samples.
The dilepton background is characterized by 
an ARGUS function.
The shape of each background is determined from a MC sample.
 (The $K^{(*)} e^{\pm} \mu^{\mp}$ background shape is found to be consistent in MC and data.)
Since the shape for $K^{(*)} \ell^{\pm} h^{\mp}$ is similar to that for the dilepton background, we use the same parameterizations for both backgrounds. 
The residual background from $\psi X_s$ is estimated from a MC sample of
$\psi$ inclusive events and parametrized by the sum of an ARGUS
function and a Gaussian. 
The background from events with misidentified leptons 
is also parametrized by the sum of an ARGUS function and a Gaussian.
In the fit, all background fractions except the dilepton background are fixed while the signal fraction is allowed to float.

\begin{figure}[htbp]
\begin{center}
  \includegraphics[scale=0.48]{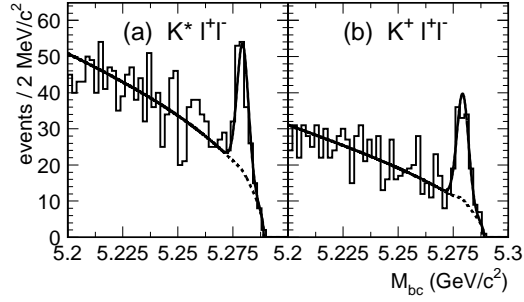}
\end{center}
\caption{$M_{\rm bc}$ distributions for (a) $B \to K^{*} \ell^{+} \ell^{-}$ 
and (b) $B\to K^+ \ell^{+} \ell^{-}$ samples. The solid
and dashed curves are the fit results for the total and background
contributions.}
\label{fig:mbcfit}
\end{figure}

Figure~\ref{fig:mbcfit} shows the fit result. We obtain $113.6 \pm 13.0$ and $96.0 \pm 12.0$ signal events for $\Kstll$ and $K^+\ell^+\ell^-$, respectively.


We use $B \to K^* \ell^+ \ell^-$ candidates in the signal box to measure the normalized double differential decay width. 
For the evaluation of the Wilson coefficients, the NNLO Wilson coefficients $\Ci$ of Ref.~\cite{WC} are used. Since the full NNLO calculation only exists for $q^2/m_b^2<0.25$ region, we adopt the so-called partial NNLO calculation~\cite{ALGH} for $q^2/m_b^2>0.25$.
The higher order terms in the $\Ci$ are fixed to the SM values while the leading terms $A_i$, with the exception of $A_7$, are allowed to float. Since the branching fraction measurement of $B \to X_s \gamma$ is consistent with the prediction within the SM, $A_7$ is fixed at the SM value, $-0.330$, or the sign-flipped value, $+0.330$. 
We choose $A_9/A_7$ and $A_{10}/A_7$ as fit parameters.
The SM values for $A_9$ and $A_{10}$ are 4.069 and -4.213, respectively~\cite{ALGH}.
To extract the these ratios, we perform an unbinned maximum likelihood fit to the events in the signal box with a probability density function (PDF) that includes the normalized double differential decay width.
The PDF used for the fit consists of terms describing the signal, cross-feeds and backgrounds:
\begin{eqnarray}
& &P(\Mbc,q^2,\cos\theta;A_9/A_7,A_{10}/A_7) \nonumber\\
                &=& \frac{1}{N_{\rm sig}}f_{\rm sig}\epsilon_{\sig}(q^2,\cos\theta)g(q^2,\cos\theta) \nonumber\\
		&+& \frac{1}{N_{\rm CF}} f_{\rm CF}\epsilon_{\rm CF}(q^2,\cos\theta)g(q^2,\cos\theta)\nonumber \\
		&+& \frac{1}{N_{\rm IF}}f_{\rm IF}\epsilon_{\rm IF}(q^2,\cos\theta)g(q^2,-\cos\theta)\nonumber  \\
		&+& (1-f_{\rm sig}-f_{\rm CF}-f_{\rm IF}-f_{\Kstarhh}-f_{\psi X_s}) \times \nonumber\\
                &&\Big\{(f_{\Kstarlh}{\cal P}_{\Kstarlh}(q^2,\cos\theta) + (1-f_{\Kstarlh}){\cal P}_{\rm dl}(q^2,\cos\theta)\Big\} \nonumber\\
		&+& f_{\Kstarhh}{\cal P}_{\Kstarhh}(q^2,\cos\theta) + f_{\psi X_s}{\cal P}_{\psi X_s}(q^2,\cos\theta).
\label{eq:pdf}
\end{eqnarray}
Here, ${\cal P}_{\Kstarlh}$, ${\cal P}_{\rm dl}$, ${\cal P}_{\Kstarhh}$ and ${\cal P}_{\psi X_s}$ are the probability density functions for 
$\Kstarlh$, dilepton background, $\Kstarhh$ and $\psi X_s$, respectively.
The quantities $\epsilon_{\rm sig}$ ($N_{\rm sig}$), $\epsilon_{\rm CF}$ ($N_{\rm CF}$) and $\epsilon_{\rm IF}$ ($N_{\rm IF}$) correspond to the efficiency function (normalization) of each signal and cross-feed component.
Each fraction $f$ is the probability of finding the corresponding component in the data sample for a given $M_{\rm bc}$ value determined from the $\Mbc$ fit, with the exception of $f_{\Kstarlh}$, which is the fraction within the dilepton background component determined from the MC samples.) 
The functions $\epsilon$ and ${\cal P}$ for the dilepton background, $\Kstlh$ and $\psi X_s$ are obtained from MC samples. 
The $\Ksthh$ background shape ${\cal P}_{\Kstarhh}$ is obtained from $\Ksthh$ events and the momentum- and angular-dependent hadron to lepton misidentification probability.

The renormalization scale $\mu$ is set to 2.5~GeV as suggested by Ref.~\cite{ALGH}.
The double differential decay width includes the form factor parameters and the bottom quark mass $m_b$. We choose the form factor model of Ali {\it et al.}~\cite{ALGH,bib:abhh} and a bottom quark mass of $4.8$~GeV$/c^2$. 

First, we measure the integrated asymmetry $\tilde{{\cal A}}_{\rm FB}$, which is defined as
\begin{eqnarray}
\tilde{{\cal A}}_{\rm FB} ={\int  \int^1_{-1} {\rm sgn}(\cos\theta)g(q^2,\theta) d\cos\theta dq^2\over \int \int^1_{-1} g(q^2,\theta) d\cos\theta dq^2}.
\end{eqnarray}
We determine the yield in each $q^2$ and forward-backward regions from a fit to the $\Mbc$ distribution. Then we correct the efficiency and obtain
\begin{eqnarray}
\tilde{{\cal A}}_{\rm FB}(B \to \Kstll) &=& 0.50 \pm 0.15 \pm 0.02, \nonumber \\
\tilde{{\cal A}}_{\rm FB}(B^+ \to K^+ \ell^+ \ell^-) &=& 0.10 \pm 0.14 \pm 0.01,
\end{eqnarray}
where the first error is statistical and the second is systematic.
A large integrated asymmetry is observed for $\Kstll$ with a significance of $3.4\sigma$.
The result for $K^+ \ell^+ \ell^-$ is consistent with zero as expected. 

\begin{figure}[htbp]
\begin{center}
  \includegraphics[scale=0.75]{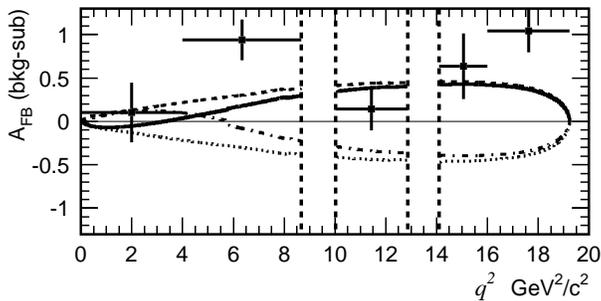}
\end{center}
\caption{Fit result for the negative $A_7$ solution (solid) projected onto the background subtracted forward-backward asymmetry, and forward-backward asymmetry curves for several input parameters, including the effects of efficiency; $A_7$ positive case~($A_7=0.330$, $A_9=4.069$, $A_{10}=-4.213$) (dashed), $A_{10}$ positive case ($A_7=-0.280$, $A_9=2.419$, $A_{10}=1.317$) (dot-dashed) and both $A_7$ and $A_{10}$ positive case~($A_7=0.280$, $A_9=2.219$, $A_{10}=3.817$) (dotted)~\cite{C10flip}. The new physics scenarios shown by the dot-dashed and dotted curves are excluded.}
\label{fig:afb}
\end{figure}

We fit the $\Kstll$ candidates with the PDF of Eq.~\ref{eq:pdf}.
The fit results of ratios of Wilson coefficients are summarized in Table~\ref{tab:result}.
Figure~\ref{fig:afb} shows the fit results projected onto the background-subtracted forward-backward asymmetry distribution in bins of $q^2$. 
\begin{table}[htbp]
  \caption{%
    $A_9/A_7$ and $A_{10}/A_7$ fit results for negative and positive $A_7$ values. The first error is statistical and the second is systematic.}
 \begin{center}
 \begin{tabular}{lcccc} \hline \hline
		 & & Negative $A_7$       & & Positive $A_7$ \\ \hline
$A_9/A_7$        &  & $-15.3 \PM{3.4}{4.8}\pm1.1$ &  & $-16.3 \PM{3.7}{5.7}\pm1.4$ \\
$A_{10}/A_7$     &  & $10.3 \PM{5.2}{3.5}\pm1.8$  &  & $11.1 \PM{6.0}{3.9}\pm2.4$ \\ \hline \hline
  \end{tabular}
   \end{center}
\label{tab:result}
\end{table}


We estimate contributions to the systematic error due to 
uncertainties in the physics parameters, finite $q^2$ resolution, 
efficiency and signal probability.
We vary the $A_7$ value within the range allowed by the branching fraction of $B \to X_s \gamma$~\cite{bib:hiller}.
The bottom quark mass $m_b$ is varied by $\pm0.2$~GeV$/c^2$. The systematic uncertainty associated with the choice of the form factor model is taken from the difference in fit results using the models of Ali {\it et al.} and Melikhov {\it et al.}~\cite{bib:mns}.
The effect of $q^2$ resolution is estimated using a toy MC study. The effect due to $\cos\theta$ resolution is found to be negligible.
The uncertainty in the efficiency is estimated by
changing the efficiency for pions with $p<0.3$~GeV$/c$, electrons with $p<0.7$~GeV$/c$ and muons with $p<1$~GeV$/c$ by 10\%, 5\% and 10\%, respectively, to obtain revised efficiency functions for signal and background PDFs.
We change the shape parameters for the signal or background probability functions $f$ and take the difference as an uncertainty in the signal fraction. The parameters are modified by $\pm1\sigma$ for signal, dilepton background and $\Ksthh$. We vary the normalization for cross-feed events and $\psi X_s$ by 100\% since we cannot determine the uncertainty from data. To assign the uncertainty in $\Kstlh$, we change the fraction $f_{\Kstarlh}$ by 20\%, which corresponds to the difference between MC and sideband events.
Table~\ref{tab:syst} summarizes the contributions to the systematic error. 
\begin{table}[htbp]
  \caption{%
    Summary of systematic errors. }
 \begin{center}
 \begin{tabular}{lcccccc} \hline \hline
Source & & \multicolumn{2}{c}{Negative $A_7$}& &\multicolumn{2}{c}{Positive $A_7$}  \\ 
                    & & $A_9/A_7$        & $A_{10}/A_7$      & & $A_9/A_7$         & $A_{10}/A_7$ \\ \hline
$A_7$ \cite{bib:hiller}        	    & & $\PM{0.2}{0.0}$ & $\pm0.0$ & & $\PM{0.1}{0.2}$   & $\PM{0.3}{0.1}$ \\
$m_b$ ($4.8\pm0.2$~GeV)              & & $\pm{0.7}$       & $\pm{0.5}$        & & $\pm 0.6$         & $\pm0.4$ \\
Model dependence    & & $\pm0.7$         & $\pm1.7$          & & $\pm1.0$          & $\pm2.2$ \\
$q^2$ resolution    & & $\pm0.3$         & $\pm0.4$          & & $\pm0.3$          & $\pm0.4$\\
Efficiency          & & $\pm0.1$        & $\pm0.0$         & & $\pm0.1$          & $\pm0.1$\\
Signal probability  & & $\PM{0.4}{0.5}$  & $\PM{0.2}{0.3}$   & & $\PM{0.4}{0.5}$   & $\pm{0.4}$\\ \hline
Total               & & $\pm{1.1}$       & $\pm{1.8}$        & & $\PM{1.3}{1.4}$   & $\PM{2.4}{2.3}$ \\ \hline \hline
  \end{tabular}
  \end{center}
\label{tab:syst}
\end{table}

\begin{figure}[htbp]
\begin{center}
  \includegraphics[scale=0.35]{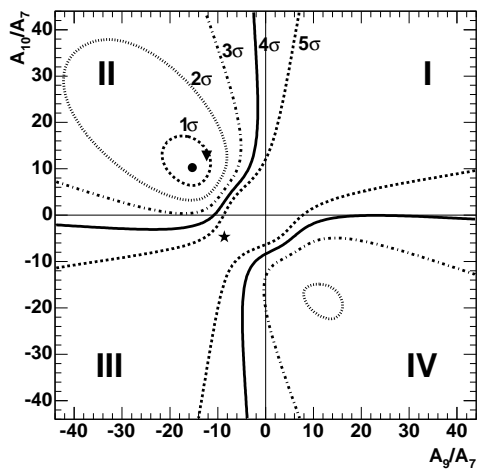}
\end{center}
\caption{CL contours for negative $A_7$. Curves show 1$\sigma$ to 5 $\sigma$ contours. The symbols show the fit (circle), SM (triangle), and $A_{10}$-positive (star)~\cite{C10flip} cases.}
\label{fig:cl}
\end{figure}

The fit results are 
consistent with the SM values ${A_9}/{A_7}=-12.3$ and ${A_{10}}/{A_7}=12.8$.
In Fig.~\ref{fig:cl}, we show confidence level (CL) contours in the ($A_9/A_7$, $A_{10}/A_7$) plane based on the fit likelihood smeared by the systematic error, which is assumed to have a Gaussian distribution.
We also calculate an interval in $A_9A_{10}/A_7^2$ at the 95\% CL for the allowed $A_7$ region,
\begin{eqnarray}
-14.0 \times 10^2 < {A_9}{A_{10}}/{A_7^2} < -26.4.
\end{eqnarray}
From this, the sign of ${A_9}{A_{10}}$ must be negative, and the solutions in quadrants I and III of Fig.~\ref{fig:cl} are excluded at 98.2\% confidence level.
Since solutions in both quadrants II and IV are allowed, we cannot determine the sign of $A_7A_{10}$.
Figure~\ref{fig:afb} shows the comparison between the fit results for the negative $A_7$ value projected onto the forward-backward asymmetry, and the forward-backward asymmetry distributions for several input parameters. We exclude the new physics scenarios shown by the dotted and dot-dashed curves, which have a positive ${A_9}{A_{10}}$ value.

%

In summary, we have measured the ratios of Wilson coefficients in $B \to \Kstll$ decay for the first time by studying the forward-backward asymmetry in the angular distribution of leptons. We observe a large integrated forward-backward asymmetry with a significance of $3.4\sigma$.
The fit results are consistent with the SM prediction and also with the case where the sign of $A_7A_{10}$ is flipped. We exclude new physics scenarios with positive $A_9A_{10}$ at 98.2\% confidence.


We would like to thank Gudrun Hiller and Enrico Lunghi for their invaluable suggestions.
We thank the KEKB group for excellent operation of the
accelerator, the KEK cryogenics group for efficient solenoid
operations, and the KEK computer group and
the NII for valuable computing and Super-SINET network
support.  We acknowledge support from MEXT and JSPS (Japan);
ARC and DEST (Australia); NSFC and KIP of CAS (contract No.~10575109 and 
IHEP-U-503, China); DST (India); the BK21 program of MOEHRD, and the
CHEP SRC and BR (grant No. R01-2005-000-10089-0) programs of
KOSEF (Korea); KBN (contract No.~2P03B 01324, Poland); MIST
(Russia); ARRS (Slovenia);  SNSF (Switzerland); NSC and MOE
(Taiwan); and DOE (USA).

%


\end{document}